\documentclass[aps,nofootinbib]{revtex4-1}
\usepackage{graphicx}
\usepackage{setspace}  
\usepackage{accents}  
\begin{document}

\title{Finite Size Scaling for Criticality of  the Schr\"odinger  Equation}

\author{Sabre Kais\footnote{kais@purdue.edu}}
\affiliation{Department of Chemistry,
Purdue University,
West Lafayette, IN 47907}

\begin{abstract}
By solving the Schr\"odinger equation  one obtains the 
whole energy spectrum, both the bound and the continuum states. If the Hamiltonian depends on a set
of parameters, these could be tuned to a transition  from bound to continuum
states.  The behavior of systems
 near the threshold, which separates bound-states from continuum states,
is important in the study of such phenomenon as: ionization of atoms and molecules, 
molecule dissociation, scattering collisions and stability of matter.
In general, the energy is non-analytic as a function of  the Hamiltonian
parameters or a bound-state does not exist
at the threshold energy. 
The overall goal of this chapter is to show how one can 
predict, generate and identify new class of stable 
quantum systems using large-dimensional models and the finite size scaling approach. Within
this approach,  the finite
size corresponds not to the spatial  dimension but to the number of elements
in a complete basis set used to expand the exact  eigenfunction
of a given Hamiltonian. This  
method is efficient and very accurate for estimating the critical 
parameters, $\{\lambda_i\}$, 
for stability of a given Hamiltonian, $H(\lambda_i)$.
We present two methods of obtaining critical parameters using finite size scaling for a given quantum
Hamiltonian:  The finite element method and the basis
set expansion method. The long term goal of developing finite size scaling 
is treating criticality from first principles at quantum phase transitions. 
In the last decade considerable
attention has concentrated on a new  class of 
phase transitions, transitions which occur at the absolute zero
of temperature. These are 
quantum phase transitions which are driven by quantum fluctuations as
a consequence of
Heisenberg's uncertainty principle.  
These new transitions are  tuned by parameters in the
Hamiltonian. Finite size scaling might be useful in predicting the quantum 
critical parameters for systems going through quantum phase transitions.
\end{abstract}

\maketitle

\section{Introduction}

Weakly bound states   represent an interesting
field of research in atomic and molecular physics. The behavior of systems
 near the binding threshold
is important in the study of ionization of atoms and molecules, 
molecule dissociation and scattering collisions.
Moreover, the stability of atomic and molecular systems  in external 
electric, magnetic and laser fields  
is of
fundamental importance in atomic and molecular physics and
has attracted considerable
experimental and theoretical attention over the past
decades\cite{science,magnetic,
Laser0,qiwei}.
A superintense laser field can change the nature of atomic and molecular 
systems and their anions; the stabilization in superstrong
fields is accompanied by splitting of the electron distribution
into distinct lobes, with locations governed by the quiver
amplitude and polarization of the laser field. This localization
markedly alters electron-nucleus interactions as well as
reduces electron-electron repulsions and hence suppresses
autoionization. In molecules, it can also enhance chemical
bonding. This localization markedly
reduces the ionization probability and can enhance chemical bonding when the laser strength
becomes sufficiently strong and can give rise to new stable multiply charged 
negative ions such as H$^{--}$, He$^-$ and H$_2^-$\cite{Laser0,Laser4,Laser5,Qi-Laser}.\\

In general, the energy is non-analytical, an analytic function is a function that is locally given by
a convergent power series, as a  function of  the Hamiltonian
parameters or a bound-state does not exist
at the threshold energy. 
It has been suggested for some time, based on 
large-dimensional models, that there are
possible analogies  between
critical phenomena and singularities of the energy
\cite{stillinger1,katriel,dudley}.\\

Phase transitions are associated with singularities of the free energy. These
singularities occur only in the thermodynamic limit\cite{yanglee1,yanglee2}
where the dimension of the system approaches infinity. However calculations
are done only on finite systems.  A Finite Size Scaling (FSS) approach is
needed in order to extrapolate results from finite systems to the
thermodynamic limit\cite{fisher}.  FSS is not only a formal way to understand
the asymptotic behavior of a system when the size tends to infinity, but a
theory that also gives us numerical
methods\cite{widom,barber,privman,cardy,nightingale1,Peter1,Peter2} capable of obtaining
accurate results for infinite systems by studying the corresponding small
systems\cite{neirotti0,serra2,kais,snk1,snk2,nsk,qicun,kais1,review,adv,dipole,quadrupole}. 
Applications  include expansion in Slater-type basis functions\cite{adv},  
Gaussian-type  basis functions\cite{gto} and recently  
finite elements\cite{fem-fss}.

\section{Criticality for  Large-Dimensional Models}

Large dimension models were originally developed for
specific theories in the
fields of nuclear physics, critical phenomena and particle physics\cite{D3,witten}.
Subsequently, with the pioneering work of Herschbach\cite{DudleyJCP,dudley},
they found wide use in the field of
atomic and molecular physics\cite{D4}.
In this method one takes
the dimension of space, $D$, as a variable, solves the problem
at some
dimension $D \not= 3$ where the physics becomes much simpler,
and then uses perturbation
theory or other techniques to obtain an approximate
result for $D=3$\cite{dudley}. \\

It is possible to describe stability and symmetry breaking of
electronic structure configurations  of atoms and molecules
as  phase transitions and critical phenomena. 
This analogy was revealed by
using dimensional scaling method and the large dimensional limit model of
electronic structure configurations\cite{D-He,D-N,D-H2,D-AB}. \\

To study the behavior of a given system near the critical
point, one has to rely on model calculations which are  simple, capture the
main physics of the problem and which  belong to the same universality class\cite{privman,cardy}.
Here we will illustrate the phase transitions and
symmetry breaking using the large dimension model.
In the
application of dimensional scaling to electronic structure, the large-D limit
reduces to a semi-classical electrostatic problem in which the electrons
are assumed to have fixed positions relative both to the nuclei and to each other
in the D-scaled space\cite{dudley}.
This configuration corresponds to the minimum
of an effective potential which includes Coulomb interactions as well
as centrifugal terms arising from the generalized D-dependence kinetic
energy. Typically, in the large-D regime the electronic structure
configuration undergoes symmetry breaking for certain ranges of nuclear
charges or molecular geometries\cite{frantz}.\\

In order to illustrate the analogy between symmetry breaking and phase transitions
we present as an example:  the 
results for the two-electron atoms in the
Hartree-Fock (HF) approximation\cite{D-He}.
In the HF approximation at the $D \rightarrow \infty$ limit,
the dimensional-scaled effective Hamiltonian
for  the two-electron atom in  an external weak electric
field $\cal E$ can be written as\cite{loeser,cabrera},

\begin{equation}
{\cal H_{\infty}} \, = \, \frac{1}{2} \left( \frac{1}{r_1^2}
\,+\,\frac{1}{r_2^2}
\right) \,-\, Z \,  \left( \frac{1}{r_1} \,+\,\frac{1}{r_2}
\right) \,+\,
\frac{1}{\left( r_1^2 + r_2^2 \right)^{1/2}} \,-\,
{\cal E} \left( r_1 - r_2 \right)
\label{11}
\end{equation}

\noindent where $r_1$ and $r_2$ are the electron-nucleus radii,
and  $Z$ is
the nuclear charge.
The ground state energy at the large-D limit is then given by
${ E}_{\infty}(Z,{\cal E}) \,=\, \min_{\{r_1,r_2\}} \;
{\cal H_{\infty}}$.

In the absence of an external electric field, ${\cal E}=0$,
Herschbach
 and coworkers\cite{goodson}
have found that these equations have
a symmetric solution with the two electrons equidistant from
the nucleus,
with $r_1=r_2=r$. This symmetric solution represents  a  minimum in the region
where all the eigenvalues of the Hessian matrix
are positive, $Z \, \ge \, Z_c \,=\,
\sqrt{2}$.
For values of $Z$ smaller than $Z_c$,
 the solutions
 become 
unsymmetrical with one electron much closer to the nucleus than the other
($r_1 \neq r_2 $). In order to describe this symmetry breaking, it is
convenient to introduce
new variables $(r, \eta)$ of the form: 
$r_1 \,=\, r ;\; r_2 \,=\, (1-\eta) r$, 
where $\eta =(r_1-r_2)/r_1 \ne 0 $ measures the deviation from the symmetric solution. \\

By studying the eigenvalues of the Hessian matrix, one finds that the
solution is  a minimum of the effective potential
 for the range, $1 \le Z \le Z_c$.
We  now turn to the question of how to describe the system near the critical
point. To answer this question,
a complete mapping between this problem and
critical phenomena  in statistical mechanics
is readily feasible with the following analogies:
\begin{center}
\begin{itemize}
\item nuclear charge $(Z)  \leftrightarrow $ temperature $(T)$
\item external electric field $({\cal E}) \leftrightarrow $ ordering field $(h)$
\item ground state energy $({E_{\infty}}(Z,{\cal E}))  \leftrightarrow
$ free energy $(f(T,h))$
\item asymmetry parameter $(\eta ) \leftrightarrow $ order parameter $(m)$
\item stability limit point $(Z_c,{\cal E}=0) \leftrightarrow
$ critical point $(T_c,h=0)$
\end{itemize}
\end{center}

Using the above  scheme, we can define the critical
exponents $(\beta, \alpha, \delta$ and $\gamma)$ for the
electronic structure of the two electron atom in the following way:

\begin{equation}
\begin{array}{llll}

\eta(Z,{\cal E}=0) & \sim
 & (- \Delta Z)^\beta & \;; \;{ \Delta Z
\rightarrow 0^-}  \\

{E_{\infty}}(Z,{\cal E}=0) & \sim
 & \mid \Delta Z \mid^{\alpha}&\; ;\;{\Delta Z
\rightarrow 0}\\

{\cal E}(Z_c,\eta) & \sim & \eta^\delta sgn(\eta)& \;;\;{\eta \rightarrow 0}\\

\frac{\partial \eta}{\partial {\cal E}}\left|_{{\cal E}=0} \right. &
\sim &
 \mid \Delta Z \mid^{-\gamma}&\;;\;\Delta Z \rightarrow 0
\end{array}
\label{61}
\end{equation}
where $\Delta Z \equiv Z - Z_c$.
These critical exponents describe the nature of the singularities
in the above quantities at the critical charge $Z_c$.
The values obtained for these critical exponents are known as classical
or mean-field critical exponents:
$\beta \,=\, \frac{1}{2} \;\;;\;\; \alpha \,=\, 2,  \;\;;\;\; \delta \,=\, 3
\;\;;\;\; \gamma \,=\, 1$. \\

This  analogy between symmetry breaking and phase transitions
was  also generalized to include the
large dimensional model of the N-electron atoms\cite{D-N}, simple diatomic
 molecules\cite{D-H2,D-AA},
both linear and planar one-electron systems\cite{D-AB} as well as three-body Coulomb
systems of the general form $ABA$\cite{D-ABA}. \\

The above simple  large-D picture helps to establish  a
connection to phase
transitions. However, the next  question to be addressed is:
How to carry out such an  analogy to  $D=3$?. 
This question  will be examined in the subsequent sections using the
finite size scaling approach.

\section{Finite Size Scaling: A Brief History}

Ice tea, boiling water and other aspects of two-phase coexistence are
familiar features of daily life. Yet phase transitions do not exist at all in finite
systems! They appear in the thermodynamic limit: The volume $V \rightarrow \infty$ and
particle number $N \rightarrow \infty$ in such a way that 
their ratio, which is the density $\rho=N/V$, approaches a finite
quantity. In statistical mechanics, the existence of
phase transitions is associated with singularities of the free
energy per particle in some region of the thermodynamic space.
These singularities occur only in the {\it thermodynamic limit}\cite{yanglee1,yanglee2}.
This fact could be  understood by examining  the partition
function $Z$.   
\begin{equation}
Z= \sum_{microstate \; \Omega} e^{-E(\Omega)/k_B T},
\end{equation}
where $E(\Omega)$ is the energies of the states, $k_B$ is the Boltzmann constant and
$T$ is the temperature.  For a finite
system, the partition function is a finite sum of analytical terms, and
therefore it is itself an analytical function. The Boltzmann factor is an 
analytical function of $T$ except at $T=0$. For $T> 0$, 
it  is necessary
to take an infinite
number of terms in order to obtain a singularity
in the thermodynamic limit\cite{yanglee1,yanglee2}.\\

In practice,  real systems have a large but finite volume and particle numbers
($N \sim 10^{23}$),
and phase transitions are observed.
More dramatic even is the case of
numerical simulations, where sometimes systems with only a few number
(hundreds, or even
tens) of particles are studied, and ``critical" phenomena are still
present. Finite size scaling theory, which  was pioneered by Fisher\cite{fisher},  
addresses
the  question of why finite systems apparently
describe  phase transitions and what is the
relation of this phenomena with the true phase transitions in 
corresponding infinite systems. Moreover,
 finite-size scaling is not only a formal way to understand the
asymptotic
behavior of a system when the size tends to infinity. In fact, the theory
gives us numerical methods  capable
of  obtaining  accurate  results for infinite systems  by studying
the corresponding small systems (see \cite{barber,privman,cardy} and
references therein).\\

In order to understand the  main idea of finite size scaling, let us consider
a system defined in a $D$-dimensional volume $V$ of a linear dimension $L$
($V=L^D$). In  a finite size system, If quantum effects are not taken into consideration,
 there are in principle three length scales: The finite geometry characteristic size L,
the correlation length $\xi$, which may be defined as the length scale covering the
exponential decay $e^{-r/{\xi}}$ with distance $r$ of the correlation
function,  and the microscopic length $a$ which governs the range of the interaction.
Thermodynamic quantities thus may depend on the dimensionless ratios
$\xi/a$ and $L/a$. The finite size scaling hypothesis assumes that,
close to the critical point, the microscopic length drops out.\\

If in the thermodynamic limit,  $L \rightarrow \infty$,
we consider that there is only one parameter  (say temperature  $T$)
in the problem and the infinite system has a second order phase transition at a critical
temperature $T_c$, a thermodynamic  quantity $G$ develops a singularity
as a function of the temperature $T$ in the form:

\begin{equation}
\label{limit}
G(T)=\lim_{L\to\infty} G_L(T)\sim\left|T-T_c
\right|^{-\rho}\;,
\end{equation}

\noindent whereas it is regular in the finite system,
$G_L(T)$ has no singularity.

When the size $L$ increases, the singularity of $G(T)$ starts to develop.
For example, if the correlation length diverges at $T_c$ as:

\begin{equation}
\label{xi}
\xi(T)=\lim_{L\to\infty} \xi_L(T)\sim\left|T-T_c
\right|^{-\nu}\;,
\end{equation}

\noindent then $\xi_L(T)$ has a maximum which becomes sharper and sharper,
then FSS ansatz assumes the existence of scaling function
$F_K$ such that:

\begin{equation}
\label{kn}
G_L(T)\sim G(T) F_K \left(\frac{L}{\xi(T)}\right)\;,
\end{equation}

\noindent where $F_K(y) \,\sim\, y^{\rho/\nu}$ for  $y \sim 0^+$.
Since the FSS ansatz, Eq. (\ref{kn}), should be valid for any quantity
which exhibits
an algebraic singularity in the bulk, we can apply it to the correlation length
$\xi$ itself. Thus the correlation length in a finite system should have the
form:

\begin{equation}
\label{xin}
\xi_L(T)\sim L \phi_{\xi}(L^{1/\nu}|T-T_c|)\;.
\end{equation}

\noindent The special significance of this result was first realized by
Nightingale \cite{nightingale}, who showed how it could be reinterpreted as a
renormalization group transformation of the infinite system. The
phenomenological renormalization (PR) equation for finite systems
of sizes $L$ and $L'$ is given by:

\begin{equation}
\label{pr1}
\frac{\xi_L(T)}{L}=\frac{\xi_{L'}(T')}{L'}\,,
\end{equation}

\noindent and has a fixed point at $T^{(L,L')}$.
It is expected that the
succession of points $\left\{T^{(L,L')}\right\}$ will
converge to the true $T_c$ in the infinite size limit.

The finite-size scaling theory combined with transfer matrix calculations
had been, since the development of
the phenomenological renormalization in 1976 by Nightingale\cite{nightingale},
one of the most powerful
tools to study critical phenomena in two-dimensional lattice models.
For these models the partition function and all the
physical quantities of  the system (free energy, correlation length,
response functions, etc)  can be written as a function of the eigenvalues of
the transfer matrix\cite{thompson}. In particular, the free energy takes the
form:

\begin{equation}
\label{fe}
f(T)=-T \ln\lambda_1
\end{equation}

\noindent and the correlation length is:

\begin{equation}
\label{cl}
\xi(T)=-\frac{1}{ \ln\left(\lambda_2/\lambda_1\right)}
\end{equation}

\noindent where $\lambda_1$ and $\lambda_2$ are the largest and the second
largest eigenvalues of the transfer matrix.
In this context, critical
points are related with the degeneracy of these eigenvalues. For finite
transfer matrix the largest
eigenvalue is isolated (non degenerated) and phase transitions can occur
only in the limit $L \rightarrow \infty$ where the size of the transfer
matrix goes to infinity and the largest eigenvalues can be degenerated.
In the next section, we will see that these ideas of finite size scaling 
can be generalize to quantum mechanics, in particular addressing the criticality
of the Schr\"odinger  equation.

\section{Finite Size Scaling for the Schr\"odinger  Equation}

The finite size scaling method is a systematic way to extract
the critical behavior of an infinite system from analysis on finite
systems\cite{adv}. It is efficient and accurate for the calculation of
critical parameters of the Schr\"odinger equation. Let's assume we
have the following Hamiltonian:
\begin{equation}
\label{h1}
{\cal H} \,=\, {\cal H}_0 \,+\, V_\lambda \;
\end{equation}

\noindent where ${\cal H}_0$ is $\lambda$-independent and $ V_\lambda$
is the $\lambda$-dependent  term. We are interested in the study of
how the different properties of the system change when the value of
$\lambda$ varies.
A critical point, $\lambda_c$,  will be  defined as a point for which a
bound state becomes absorbed or degenerate with a continuum. \\
 
Without loss of generality, 
we will assume that the Hamiltonian, Eq. (\ref{h1}),  has a bound
state, $E_\lambda$,
for
$\lambda > \lambda_c $ which becomes equal to zero at
$\lambda = \lambda_c$. As in statistical mechanics,
we can define some critical exponents related
to the asymptotic behavior of different quantities near the critical point.
In particular, for the energy we can define the critical exponent $\alpha$
as:

\begin{equation}
\label{alphaiv}
E_\lambda \, \hbox{}_{\stackrel{\hbox{\normalsize$\sim$}}
{\hbox{\scriptsize$\lambda\to\lambda_c^+$}}} (\lambda - \lambda_c)^\alpha.
\end{equation}

The existence or absence of a bound state at the
critical point is related to the type of the singularity in the energy.  
Using statistical mechanics terminology, we can associate
``first order phase transitions" with the existence of a normalizable
eigenfunction at the critical point. The absence of such a function 
could be related to  ``continuous phase transitions"\cite{adv}. \\

In quantum calculations, the  variational  method is widely used to
approximate the solution of the Schr\"odinger   equation. To  obtain
exact results one should expand the exact wave function in a
complete basis set and take the number of basis functions to
infinity. In practice, one truncates this expansion at some
order $N$.  In
the present approach,
the finite
size corresponds not to the spatial  dimension, as in statistical
mechanics,  but to the number of elements
in a complete basis set used to expand the exact  eigenfunction
of a given Hamiltonian.
We will compare  two methods to obtain the matrix elements needed to apply
the FSS ansatz. The size of our system for the basis set expansion will
correspond to the dimension of the Hilbert space. For a given
complete basis set {$\Phi_n$}, the ground-state eigenfunction has
the following expansion:
\begin{equation}
\Psi_\lambda= \sum_{n}^{}a_n(\lambda)\psi_n,
\end{equation}
where n is the set of quantum numbers. We have to truncate the
series at order N and the expectation value of any general operator $O$ at
order N is given by:
\begin{equation}
\left<O\right>^{N} = \sum_{n,m}^N a_n^{(N)}a_m^{(N)}O_{n,m},
\end{equation}
\noindent where ${\cal O}_{n,m}$ are the matrix elements of ${\cal O}$
in the basis set $\{\psi_n\}$. \\

For the finite element method (FEM), the wavefunction  
$\psi_n(r)$ in the $n$-th element is expressed in terms of local shape
functions. For our calculations, we use Hermite interpolation polynomials
with two nodes and three degrees of freedom.  
This choice ensures the continuity of the wavefunction and its
first two derivatives.
Then in $n$-th element the wavefunction is\cite{fem-fss}:
\begin{equation}
  \psi_n(r)= \sum_{i=1}^2 \left [ 
    \phi_{i}(r) \psi_n^{i} +
    \bar{\phi}_{i}(r) {\psi_n^{'i}} +
    \accentset{=}{\phi}_{i}(r) {\psi_n^{''i}} \right],
  \label{localpsi}
\end{equation}
with $\alpha$ indicating the nodal index of the element; $i=1$ for the
left and $i=2$ for the right border of the element. The functions
$\phi_{i}(r)$, $\bar{\phi_{i}}(r)$, and
$\accentset{=}{\phi}_{i}(r)$ are the (fifth degree) Hermite interpolation
polynomials.
Then $\psi_n^{i}$, $\psi_n^{'i}$, and
$\psi_n^{''i}$ are the undetermined values values of the wavefunction
and its first and second derivative on the nodal points.
The size for the case of solving the equation with the FEM will be the number
of elements used. \\

%%%%%%%%%%%%%%%%%%%%%%%%%%%%%%%%%%%%%%%%%%%
Since $\left< O \right>_\lambda$ is not analytical at
$\lambda=\lambda_c$, then we define a critical exponent, $\mu_O$, if the general operator has
the following relation:
\begin{equation}
\left< O \right>_\lambda \approx (\lambda - \lambda_c)^{\mu_O} \, \,
\, \, {for} \,\,\,\, \lambda \rightarrow \lambda_c^+,
\end{equation}
where $\lambda \rightarrow \lambda_c^+$ represents taking the limit
of $\lambda$ approaching the critical point from larger values of
$\lambda$. As in the FSS ansatz in statistical mechanics \cite{privman,derrida1}, we will assume 
that there exists a scaling function for the truncated magnitudes
such that:
\begin{equation}
\left< O \right>_\lambda^{(N)} \sim  \left< O \right>_\lambda F_O
(N|\lambda-\lambda_c|^\nu),
\end{equation}
with the scaling function $F_O$ being particular for different
operators but all having the same unique scaling exponent $\nu$.

To obtain the critical parameters, we define the following function:
\begin{equation}
\triangle_O(\lambda;N,N')=\frac{\ln(\left<O\right>_
\lambda^{N}/\left<O\right>_\lambda^{N'})} {\ln(N'/N)}.
\label{fourteen}
\end{equation}
At the critical point, the expectation value is related to $N$ as a
power law, $\left<O\right> \sim N^{\mu_O/\nu}$, and Eq.
(\ref{fourteen}) becomes independent of $N$. For the energy operator
$O=H$ and using the critical exponent $\alpha$  for the
corresponding exponent $\mu_O$ we have:
\begin{equation}
\triangle_H(\lambda_c;N,N')=\frac{\alpha}{\nu}.
\label{alphanu}
\end{equation}

In order to obtain the critical exponent $\alpha$ from numerical
calculations, it is convenient to define a new function\cite{adv}:
\begin{equation}
\Gamma_\alpha(\lambda,N,N')=\frac{\triangle_H(\lambda;N,N')}{\triangle_H(\lambda;N,N')-\triangle_{\frac{\partial
      V_\lambda}{\partial \lambda}}(\lambda;N,N')},
\label{gammafunc}
\end{equation}
which at the critical point is independent of $N$ and $N'$ and takes
the value of $\alpha$. Namely, for $\lambda=\lambda_c$ and any
values of $N$ and $N'$ we have
\begin{equation}
\Gamma_\alpha(\lambda_c,N,N')=\alpha,
\end{equation}
and the critical exponent $\nu$ is readily given by Eq. (\ref{alphanu}).

\section{The Hulthen Potential}

To illustrate the application of the FSS method in quantum mechanics, 
let us give an example of the criticality of the Hulthen potential.
The Hulthen potential behaves like a Coulomb potential for
small distances whereas for large distances it decreases exponentially so that the ``capacity''
for bound states is smaller than that of Coulomb potential. Thus, they have the same singularity
but shifted  energy levels. They always lie lower in the Coulomb case than in the Hulthen case,
where there remains only space for a finite number of bound states\cite{flugge}.
Here, we present the FSS calculations using two  methods: finite elements  and basis set 
expansion; each used
to obtaining quantum critical parameters for the Hulthen Hamiltonian. 
First, we give the analytical solution, then FSS with 
basis set expansion and finite element solution. 

%%%%%%%%%%%%%%%%%%%%%

\subsection{Analytical Solution}
The Hulthen potential has the following form\cite{hulthen,flugge}:
\begin{equation}
 V(r) = - \frac{\lambda}{a^2} \frac{{e}^{-r/a}}{1-{e}^{-r/a}}
\end{equation}
where $\lambda$ is the coupling constant, and $a$ is the scaling parameter.
For small values of $r/a$ the potential $V(r) \rightarrow  - \frac{1}{a}\lambda/r$, 
whereas for large values of $r/a$ the potential approaches zero
exponentially fast, therefore the \emph{scale a} in the potential regulates 
the infinite number of levels that would otherwise appear with a
large-distance \emph{Coulomb} behavior.

Shr\"odinger radial differential equation in the dimensionless variable $r=r/a$ becomes:
\begin{equation}
\frac{1}{2}\frac{d^2 \chi}{d r^2} + ( - \alpha^2 + \lambda 
\frac{{e}^{-r}}{1-{e}^{-r}} ) \chi = 0.
\end{equation}
We only consider the case for $l=0$ for the Hulthen potential.
Here we used the abbreviations $\alpha^2 = - E a^2 \geq 0 $ (in atomic units $m=\hbar=1$).
The complete solutions for the wavefunctions are written in term of 
hypergeometric functions as follows\cite{flugge}:

\begin{equation}
\chi = N_0 e^{-\alpha r}(1-e^{-r}) _{2}F_{1} (2\alpha+1+n,1-n,2\alpha+1;e^{-r}),
\label{hulthenWaveFnc}
\end{equation}

where the normalization factor is given by
$N_0=[\alpha(\alpha+n)(2 \alpha+n)]^{\frac{1}{2}}[\Gamma(2\alpha+n)/\Gamma(2\alpha+1)\Gamma(n)]$.
It follows that the energy levels are given by:
\begin{equation}
E_{n} = - \frac{1}{a^2} \frac{(2\lambda-n^2)^2}{8 n^2}; \,\,\, n=1,2,3...,n_{max}.
\end{equation}

We can make the following comments concerning the energy levels 
obtained for the Hulthen potential. There exists a \emph{critical} 
value for the coupling $\lambda_{c}$ to have the
  given energy levels, $\lambda_{c} = n^2 / 2$. 
  It follows directly from the first observation that the number of 
  levels $n_{max}$ allowed is \emph{finite} and it
  depends on the size of the coupling constant $n^2_{max} \leq 2 \lambda$.
As $\lambda \rightarrow \infty$ the potential is well behaved, which  
can be  seen as follows: In this limit we get the obvious inequality 
$ \alpha^2 \ll 2\lambda \Rightarrow \sqrt{2\lambda} \approx n$.
It follow that we can set $ \alpha \approx 0$ in Eq.  (\ref{hulthenWaveFnc}) to obtain:
\begin{equation}
\chi_{\alpha \rightarrow 0} = (1-e^{-r}) _{2}F_{1} (1+n,1-n,1;e^{-r}),
\end{equation}
which is the wave function at threshold. This wave function is not normalizable as
expected when the energy exponent $\alpha=2$, $E \sim (\lambda-\lambda_c)^{\alpha}$.
For the ground state, the asymptotic limit of the probability density  
for $r>>1$ and $\lambda \rightarrow \lambda_c$
becomes:
\begin{equation}
P(r) \sim  e^{-r/\xi},\;\;  \xi \sim |\lambda-\lambda_c|^{-\nu},
\end{equation}
with a characteristic length $\xi$ and exponent $\nu=1$. 
The Hulthen potential has a finite capacity determined by the critical
coupling, $\lambda_{c}$. The potential admits bound states between the range 
of values for the coupling: $\lambda = [1/2, \infty)$.

\begin{figure}[htp]
\begin{center}
\includegraphics*[width=300pt]{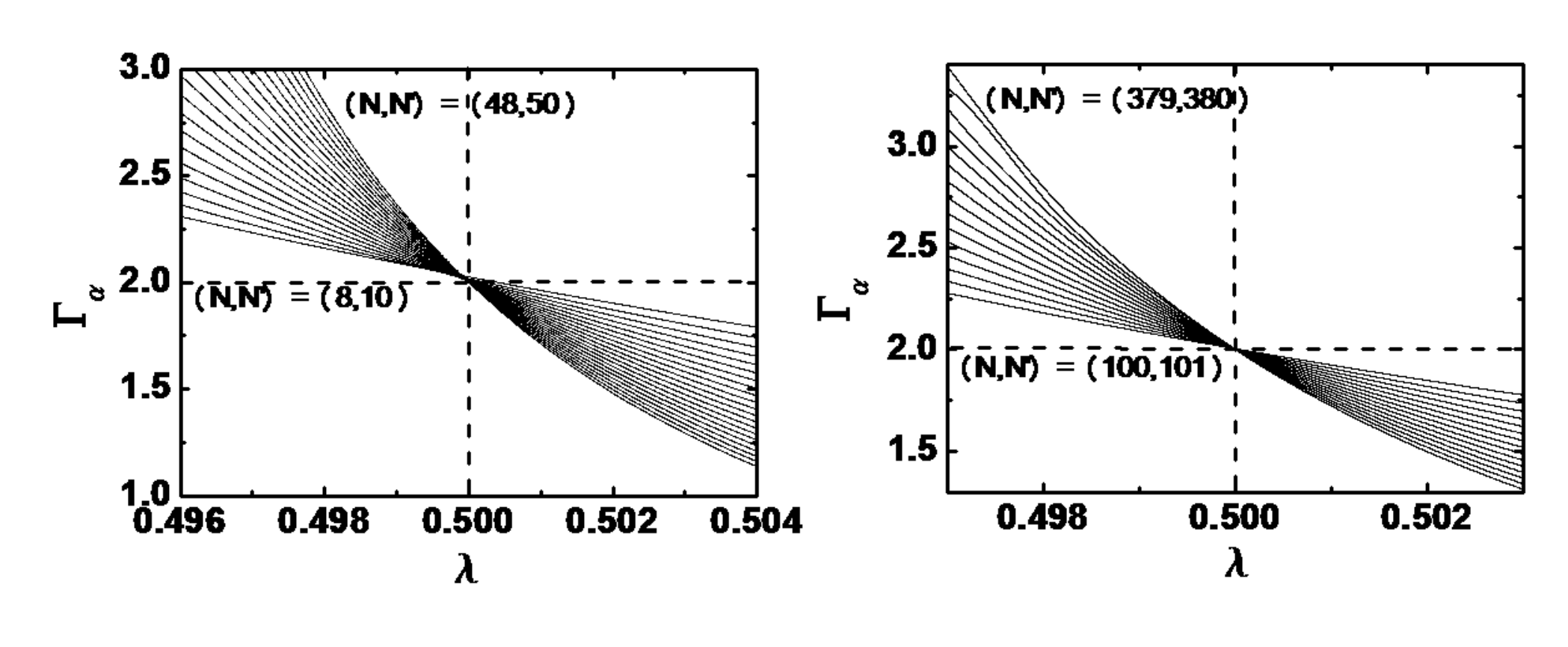}
\caption{Plot of $\Gamma_\alpha$, obtained by FSS method, as a function of $\lambda$.
  Using the number of basis N from 8 to 48 in steps of 2. For FEM the number
  elements used were from 100 to 380 in steps of 20.}
\label{gamma}
\end{center}
\end{figure}

\subsection{Basis Set Expansion}

For the Hulthen potential, the wavefunction can be expanded
in the following Slater basis ( see Chapter 7 for details \cite{chapter7}):
\begin{equation}
\Phi_n(r)=\sqrt{\frac{1/4\pi}{(n+1)(n+2)}}e^{-r/2}L_n^{(2)}(r).
\end{equation}
$L_n^{(2)}(r)$ is the Laguerre polynomial of degree $n$ and order $2$. 
The kinetic term can be obtained  analytically. However, the  
potential term need to  be calculated numerically\cite{Edwin}.

Figure {1}, show the results for the 
plot $\Gamma_\alpha(\lambda,N,N')$  as a function of
$\lambda$ with different $N$ and $N'$, all the curves will cross
exactly at the critical point.

\subsection{Finite Element Method}

The FEM is a numerical technique which  gives approximate solutions to
differential equations. In the case of quantum mechanics, the
differential equation is formulated as a boundary value problem\cite{pepper,reddy}. For
our purposes, we are interested in solving the time-independent
Shr\"odinger equation with finite elements.
We will require our boundary conditions to be restricted to the
Dirichlet type. For this problem, we will use two interpolation methods:
linear interpolation and Hermite Interpolation polynomials to solve for this
potential.

We  start by integration by parts and impose the boundary conditions 
for the kinetic energy and reduce it to the weak form\cite{fem-fss}:
\begin{eqnarray}
  &&\frac{1}{2}\int_0^{\infty}  r^2 {\psi^*}'(r)\psi'(r)  dr.
    \nonumber \\
  &&  \,\,\,\,\,\,\,\,\,\,\,\,\,\,\,\,\,\,\,\,\,\,\,
  \label{weak}
\end{eqnarray}

For the potential energy:

\begin{eqnarray}
  &&\int_0^{\infty} r^2 {\psi^*}(r)\psi(r) \lambda \frac{-e^{-r}}{1-e^{-r}} dr.
    \nonumber \\
  &&  \,\,\,\,\,\,\,\,\,\,\,\,\,\,\,\,\,\,\,\,\,\,\,
  \label{hulthen}
\end{eqnarray}

We calculated the local matrix elements of the potential energy by using a four point Gaussian
Quadrature to evaluate the integral. We set the cutoff for the integration to
$r_c$. To include the integration to infinity, we added  an infinite
element approximation. To do so,  we
approximate the solution of the wave function in the region of $[r_c,\infty)$ to be an exponentially decaying
function with the form  $\psi(r)=\psi(r_c)\,e^{-r}$. 

The local matrices are then 
assembled to form the complete solution and by invoking the variational
principle on  the nodal values $\psi_i$ we obtain a
generalized eigenvalue problem representing the initial Schr\"odinger 
equation:
\begin{equation}
H_{ij}|\psi_j \rangle=\epsilon U_{ij}|\psi_j \rangle.
\label{gev}
\end{equation}
Solution of Eq. (\ref{gev}) is achieved using standard numerical methods
(see Chapter 10 for details \cite{chapter10}).

\begin{figure}[htp]
\begin{center}
   \includegraphics*[width=300pt]{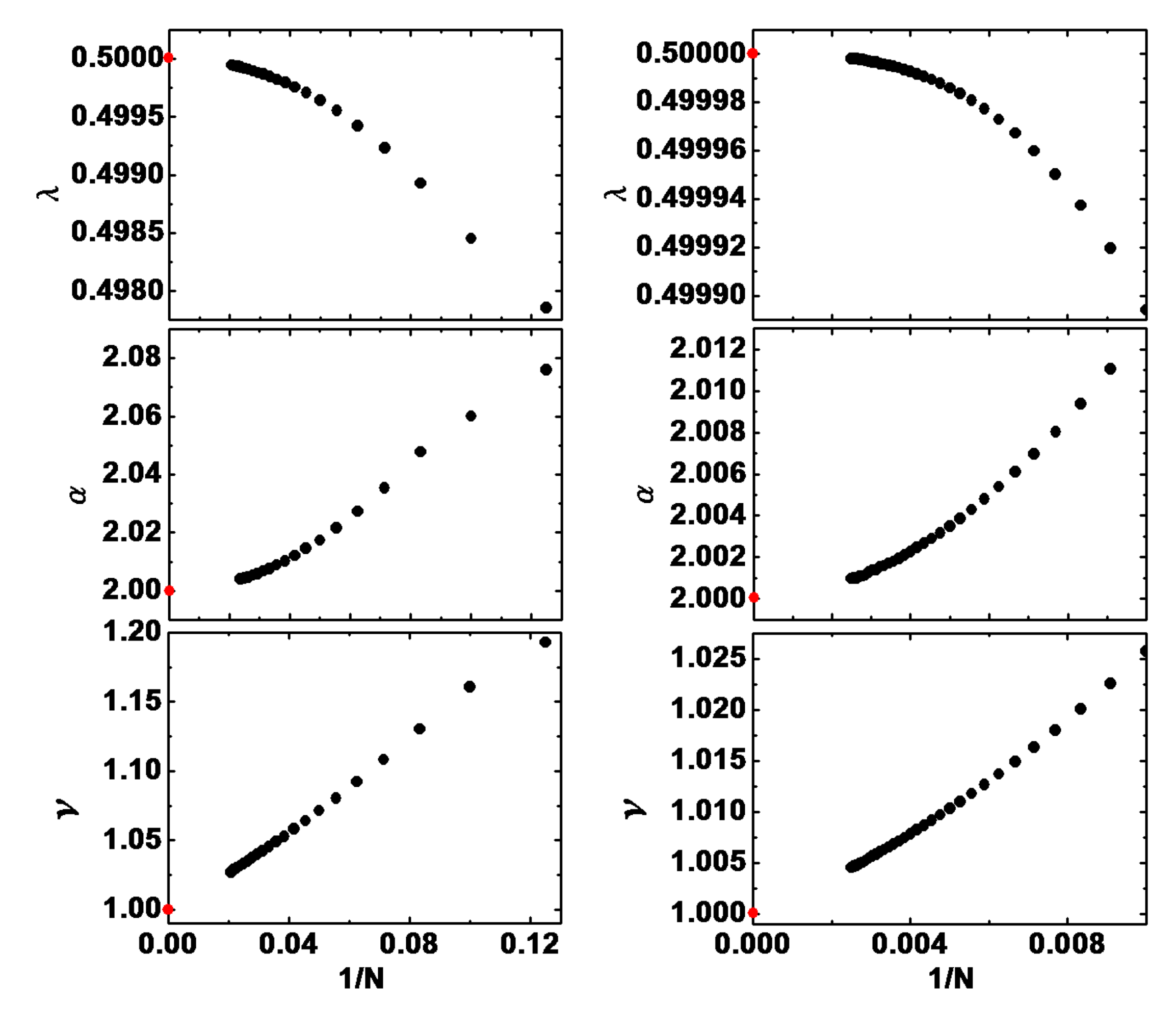}
  \caption{Extrapolated values for the critical exponents and the critical
    parameter $\lambda$. The solid read dots at $1/N=0$ are the extrapolated critical
    values. The left side is the basis set method while the right is the FEM
    with Hermite interpolation polynomials.} 
  \label{bst}
\end{center}
\end{figure}

\subsection{Finite Size Scaling Results}

\begin{figure}[htp]
\begin{center}
   \includegraphics*[width=300pt]{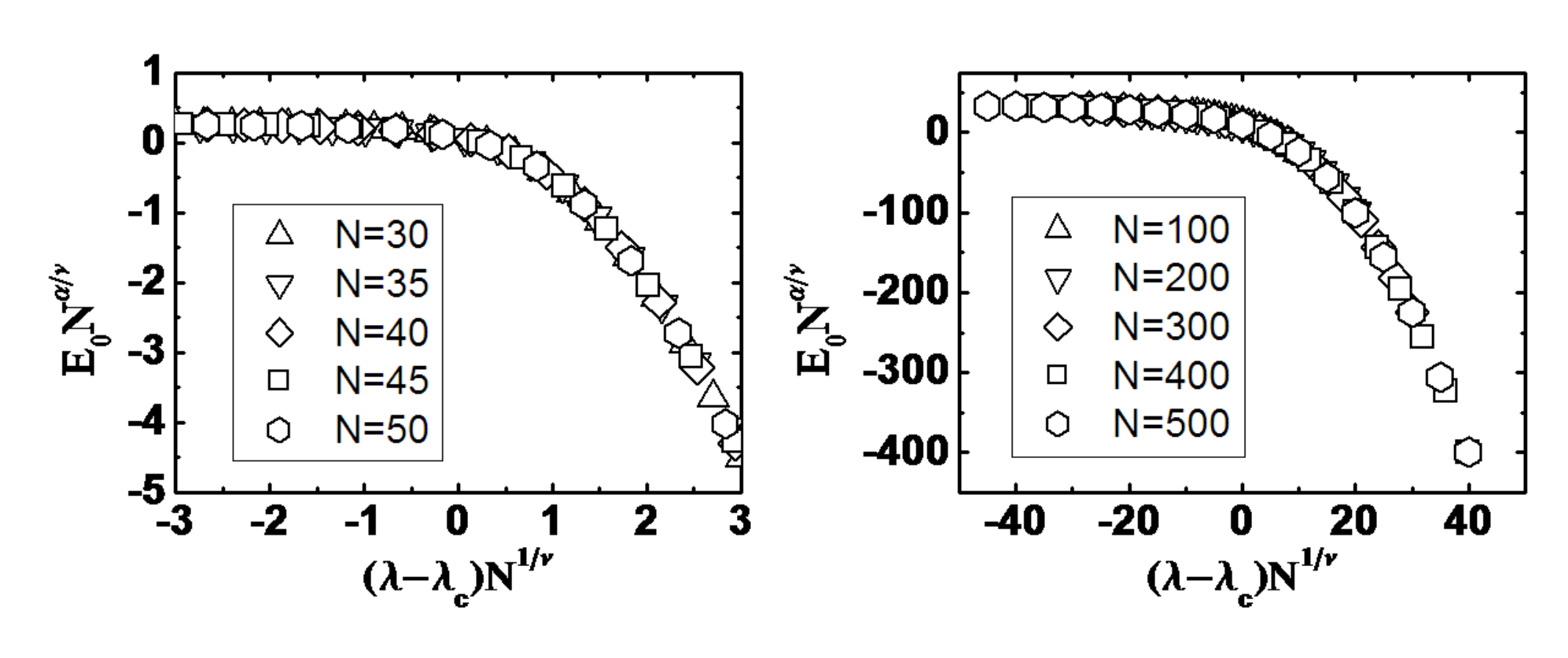}
  \caption{Data collapse study of the basis set method and FEM. The left is
    the basis set method and the right being the FEM.}
  \label{collapse}
           \end{center}
\end{figure}

The finite size scaling equations are valid only as 
asymptotic expressions, but unique values of $\lambda_c$, $\alpha$, and $\nu$
can be obtained as a succession of values as a function of $N$. The lengths of
the elements are set $h=0.5$.
The plots of $\Gamma_\alpha$,  figure
\ref{gamma},  the basis set expansion is giving values very close to
the analytical solution of the Hulthen potential. For the plot of
$\Gamma_\alpha$  for the FEM
estimation of $\lambda_c$ is producing results very close to the exact values using
Hermite interpolation. The
intersection of these curves indicate the $\lambda_c$ on the abscissa. The
ordinate gives the critical exponents $\alpha$ (in $\Gamma_\alpha$ plots). 
In Figure \ref{bst}, we observed 
the behavior of the pseudocritical parameters, $\lambda_c^{(N)}, \alpha_c^{(N)}, \nu_c^{(N)}$,
 as a function of $1/N$. 
The three curves monotonically 
converge to limiting values for the Hermite interpolation and the basis set expansion. \\

To check the validity of our finite size scaling assumptions, we
performed a data collapse\cite{datta-collapse} calculation of the Hulthen potential.  
In the  data collapse analysis, we 
examine the main assumption we have made in Eq. (17) for the existence
of a scaling function for each truncated magnitude 
$\left<{\cal O } \right>^{(N)}_\lambda $ with  a unique  
scaling exponent $\nu$.

Since the
$\left<{\cal O } \right>^{(N)}_\lambda $ is analytical in $\lambda$, 
then  from Eq. (17)  the asymptotic
behavior of the scaling function must have the form:

\begin{equation}
\label{ab}
F_{\cal O} (x) \, \sim \, x^{-\mu_{\cal O}/\nu} \;.
\end{equation}

For large values of $N$, at the $\lambda_c$, we have
\begin{equation} 
\left< {\cal O } \right>^{(N)}(\lambda_c) \sim N^{-\mu_{\cal O}/\nu}.
\end{equation}
Because the same argument of regularity holds for the derivatives of the
truncated expectation values, we have: 

\begin{equation}
\label{do}
\left. \frac{\partial^m \left<{\cal O } \right>^{(N)}}
{\partial \lambda^m} \right|_{\lambda=\lambda_c}
\sim N^{-(\mu_{\cal O}-m)/\nu}, 
\end{equation}

$\left<{\cal O } \right>^{(N)}$ is analytical in $\lambda$, then using
Eq. (\ref{do}), the Taylor expansion could be written as:

\begin{equation}
\label{te}
\left<{\cal O } \right>^{(N)}(\lambda)\sim N^{-\mu_{\cal O}/\nu}
G_{\cal O} (N^{1/\nu} (\lambda - \lambda_c)),
\end{equation}

\noindent where $G_{\cal O}$ is an analytical function of its argument.
This equivalent expression for the scaling of a given expectation value
has a correct form to study the data collapse in order to test FSS hypothesis.
 If  the scaling Eq. (17) or Eq. (\ref{te}) holds, 
then near the critical point the
physical quantities will collapse to a single universal curve when plotted 
in the appropriate form $\left<{\cal O } \right>^{(N)} N^{\mu_{\cal O}/\nu}$
against $N^{1/\nu} (\lambda - \lambda_c)$. If the operator ${\cal O}$ is the Hamiltonian
then we will have a data collapse when plotting
$E_0 N^{-\alpha/\nu}$
against $N^{1/\nu} (\lambda - \lambda_c)$.
In Figure \ref{collapse} we plot the
results corresponding to the basis set method (right panel) and Hermite interpolation 
(left panel), 
which have been calculated
with $\lambda_c=0.49999$, $\alpha=1.9960$ and $\nu=0.99910$ for the basis set method 
and for the Hermite interpolation we have $\lambda_c=0.50000$, $\alpha=2.00011$ and 
$\nu=1.000322$. The data collapse study do in fact support our FSS assumptions.
We have conveniently summarized our results for the critical parameters for
the analytical, linear interpolation, Hermite interpolation and the
basis set expansion in table \ref{table:results}.\\

\begin{table}[htp]
\begin{center}
\caption{Critical Parameters for the Hulthen Potential}
\centering
\begin{tabular}{c c c c c}
  \hline\hline
  $    $&Analytical& Linear & Hermite & Basis Set\\
  \hline
  $\lambda$ &0.5 (exact) & 0.50184 & 0.50000 & 0.49999\\
  $\alpha$ &2 (exact) & 1.99993& 2.00011 &1.9960 \\
  $\nu$ &1 (exact)& 1.00079& 1.00032&   0.99910\\ 
  \hline\hline
\end{tabular}
\label{table:results}
\end{center}
\end{table}

We have successfully obtained the critical exponents and the critical
parameter for the Hulthen potential using FSS with the basis set method and the
FEM. The results are in excellent agreement with the analytical solution even for the
very simplistic linear interpolation used for the FEM calculations. However,
the ability of the FEM to describe the wavefunction locally in terms of
elements affords a very natural way to extend its use for FSS purposes. 

\section{Finite Size Scaling and Criticality of M-Electron Atoms}

Let us examine the criticality of the N-electrons  atomic Hamiltonian as a function 
of the nuclear charge $Z$. The scaled Hamiltonian takes the form:
\begin{equation}
\label{ham}
{\cal H}(\lambda)= \sum_{i=1}^M \left[ -\frac{1}{2} \nabla_i^2 -\frac{1}{r_i} 
\right ] + \lambda \sum_{i<j=1}^M \frac{1}{r_{ij}}, 
\end{equation}

\noindent where $r_{ij}$ are the interelectron distances, and $\lambda=1/Z $ is 
the inverse of the  nuclear charge. For this  Hamiltonian, a critical point 
means the value of the 
parameter, $\lambda_c$, for
which a bound state energy becomes absorbed or degenerate
with the continuum.

To carry out the FSS procedure,  one has to choose a convenient basis set
to obtain the two lowest eigenvalues and eigenvectors of the finite 
Hamiltonian matrix. For $M=2$, one can choose the following
basis set functions:

\begin{equation}
\label{hylleraas}
\Phi_{ijk,\ell}(\vec{x}_1,\vec{x}_2) = \frac1{\sqrt2}\left(r_1^i \,r_2^j\,
e^{-(\gamma r_1+\delta r_2)} + \right.\nonumber
\left. r_1^j \,r_2^i\,e^{-(\delta r_1 + \gamma r_2)}\right)
\,\, r_{12}^k \,\;F_{\ell}(\theta_{12},{\bf\Omega})
\end{equation}
\noindent where $\gamma$ and $\delta$ are fixed parameters, 
we have found numerically that 
$\gamma = 2$ and $\delta = 0.15$ is a good choice for the ground state\cite{neirotti0},
$r_{12}$ is the interelectronic distance and $F_{\ell}(\theta_{12},{\bf\Omega})$
is a suitable function of the angle between the positions of the two electrons
$\theta_{12}$ and the Euler angles ${\bf\Omega}=(\Theta,\Phi,\Psi)$. This
function $F_{\ell}$ is different for each orbital-block of the Hamiltonian. For
the ground state $F_0 (\theta_{12},{\bf\Omega})= 1$ and
$F_1(\theta_{12},{\bf\Omega})=
\sin(\theta_{12})\cos(\Theta)$ for the $2p^2\; {}^3P$ state. 
These basis sets are complete
for each $\ell$-subspace.
The complete
wave function is then a linear combination of these terms multiplied
by variational coefficients determined by matrix diagonalization \cite{neirotti0}.
In the truncated basis set at order $N$, all terms are included such that
$N\geq i+j+k$. Using FSS calculations with $N=6,7,8,\dots,13$ gives the 
extrapolated values of $\lambda_c=1.0976 \pm 0.0004$ 
which is in excellent agreement with the best estimate of 
$\lambda_c=1.09766079$ using large-order perturbation calculations\cite{ivanov}.
Since the critical charge $Z_c=1/\lambda_c \sim 0.91 $ indicates that the 
hydrogen anion H$^{-}$ is stable, $Z=1> Z_c$.

For three-electron atoms, $M=3$, one can repeat the FSS procedure with the following
Hyllerass-type basis set\cite{serra2}:
\begin{equation}
\label{wfyd}
\Psi_{ijklmn}(\vec{x}_1,\vec{x}_2,\vec{x}_3) = {\cal C \; A}\left(r_1^i \,r_2^j
\, r_3^k r_{12}^l \, r_{23}^m \, r_{31}^n \;
e^{-\alpha (r_1+ r_2)}  e^{-\beta r_3} \;\;\chi_1
\right), 
\end{equation}

\noindent where the variational parameters,$\alpha=0.9$ and $\beta=0.1$,
 were chosen to 
obtain accurate results near the critical charge $Z \simeq 2$,
$\chi_1$ is the spin function
with spin angular moment 1/2:

\begin{equation}
\label{spin}
\chi_1 \,=\, \alpha(1) \beta(2) \alpha(3) \,-\, \beta(1) \alpha(2) \alpha(3),
\end{equation}

\noindent $ {\cal C}$ is a normalization constant and
${\cal A}$ is the usual three-particle antisymmetrizer operator\cite{serra2}.
The FSS calculations gives $\lambda_c=0.48 \pm 0.03$. Since $Z_c \sim 2.08$
the anions  He$^{-}$  and H$^{--}$ are unstable. \\

One can extend this analysis and calculate  the  critical charges 
for M-electron atoms in order to perform a systematic check of the stability of atomic
dianions. In order to have a stable doubly negatively charged atomic
ion one should require the surcharge, $S_e(N) \equiv N-Z_c(N) \geq 2$.
We have found that the
surcharge never exceeds two. 
The maximal surcharge, $S_e(86)=1.48$, is
found for the closed-shell configuration of element Rn and can be
related to the peak of electron affinity of the element $N=85$.
The FSS  numerical results for M-electron atoms  show that
at most, only one electron
can be added to a free atom in the gas phase.
The second extra electron is not
bound by singly charged negative ion because the combined action of 
the repulsive potential surrounding the isolated negative ion and
the Pauli exclusion principle. However, doubly charged 
atomic negative ions might exist in
a strong magnetic field of the order few atomic units, where
$1 a.u.=2.3505\; 10^9 G$ and superintese laser fields.

\section{Conclusions}

In this chapter, we show how the finite size scaling 
ansatz can be  combined with the variational
method  to extract information about critical behavior of 
quantum Hamiltonians. This approach is based on 
taking the number of elements
in a complete basis set or the finite element method
as the size of the system.
As in statistical mechanics,  finite size scaling
can then be used directly to the Schr\"odinger  equation.
This approach is general and gives very accurate results for 
the critical parameters, for which the bound state energy 
becomes absorbed or degenerate with a continuum.
To illustrate the applications in quantum calculations,
we present detailed calculations for the simple case of Hulthen potential
and few electron atoms. For atomic systems we have shown that 
finite size scaling can be used to explain and predict the stability 
of atomic anions: At most, only one electron can be added to a free
atom in the gas phase.\\

Recently, there has
been an ongoing experimental and theoretical search for doubly
charged negative molecular dianions\cite{science}.
In contrast to atoms, large molecular
systems can hold many extra electrons because the extra electrons
can stay well separated.
However, such systems are challenging from both
theoretical and experimental points of view.
The present finite size scaling  approach might be useful
in predicting the general stability of molecular dianions. \\

The approach can be generalize to complex systems by calculating the matrix
elements needed for FSS analysis by ab initio, density functional methods,
orbital free density functional 
(OF-DFT) \cite{princetorn,gavini}
approach, density matrices\cite{David1,David2} and other 
electronic structure methods\cite{ortiz1}.
 The implementation should be straightforward. We need to  obtain the matrix elements
to calculate $\Gamma_a$ as a function of the number of 
elements used in solving for the
system. In the finite element using mean field equations
(like Hartree-Fock or Kohn Sham methods) the solution region will be discretized into
elements composed of tetrahedrons.  \\

The field of quantum critical phenomena in atomic and molecular physics is
still in its infancy and there are many open questions about
the interpretations of the results including whether or not these quantum
phase transitions really do exist.
The possibility of exploring these 
phenomena experimentally in the field of 
quantum dots\cite{Wang-Kais} and systems in superintense laser
fields\cite{Qi-Herschbach}  offers an exciting
challenge for future research.
This finite size scaling  approach is general and 
might provide a powerful way
in determining critical parameters for the stability of atomic and molecular
systems in external fields, and for design and control electronic 
properties of materials using 
artificial atoms. \\

The critical exponents calculated with finite size scaling indicate the nature
of the transitions from bound to continuum states. Study of the 
analytical behavior of the energy near the critical point show
that the open shell system, such as the lithium like atoms, is
completely different from that of a closed shell system, such as 
the helium like atoms. The transition in the closed shell 
systems from a bound state to a continuum resemble a ``first-order
phase transition", $E \sim (\lambda - \lambda_c)^1$, 
while for the open shell system, the transition
of the valence electron to the continuum is a ``continuous  phase
transition", $E \sim (\lambda - \lambda_c)^2$. For closed shell systems,
one can show  that
${\cal H}(\lambda_c)$ has a square-integrable eigenfunction corresponding
to a threshold energy, the existence of a 
bound state at the critical coupling constant $\lambda_c$ implies that for
$\lambda < \lambda_c$, $E(\lambda)$ approaches $E(\lambda_c)$
linearly in $(\lambda-\lambda_c)$ as $\lambda \rightarrow \lambda_c^-$.
However, for open shell systems, the wave function is not 
square-integrable at at $\lambda_c$. This difference in critical exponents 
might be helpful in developing a new atomic classification schemes  based 
on the type of phase transitions and criticality of the system.

\section{Acknowledgments}
I would like to thank Pablo Serra, Juan Pablo Neirotti,
Marcelo Carignano,   Winton
Moy and Qi Wei for their valuable contributions 
to this ongoing research  of developing and applying finite size scaling to quantum 
problems and Ross Hoehn for critical reading of the Chapter. I would like also to thank the Army Research Office (ARO) 
 for financial support of this project.

\end{document}